\documentclass[prl,twocolumn,groupedaddress]{revtex4}
\usepackage[dvips]{color}
\usepackage{bm}
\usepackage{graphicx,color}
\begin{document}

\title{Selective Equal-Spin Andreev Reflections Induced by Majorana Fermions}

\author{James J. He$^1$, T. K. Ng$^1$, Patrick A. Lee$^2$}
\author{ K. T. Law$^1$} \thanks{phlaw@ust.hk}

\affiliation{$^1$ Department of Physics, Hong Kong University of Science and Technology, Clear Water Bay, Hong Kong, China \\
$^2$ Department of Physics, Massachusetts Institute of Technology, Cambridge MA 02139, USA}

\begin{abstract} 
In this work, we find that Majorana fermions induce selective equal spin Andreev reflections (SESARs), in which incoming electrons with certain spin polarization in the lead are reflected as counter-propagating holes with the same spin. The spin polarization direction of the electrons of this Andreev reflected channel is selected by the Majorana fermions. Moreover, electrons with opposite spin polarization are always reflected as electrons with unchanged spin. As a result, the charge current in the lead is spin-polarized. Therefore, a topological superconductor which supports Majorana fermions can be used as a novel device to create fully spin-polarized currents in paramagnetic leads. We point out that SESARs can also be used to detect Majorana fermions in topological superconductors.
\end{abstract}


\maketitle

\emph{\bf Introduction}--- A Majorana fermion (MF) [\onlinecite{Wilczek, Kitaev1}] is an anti-particle of itself. Due to this self-Hermitian property, MFs lead to several interesting phenomena such as fractional Josephson effects [\onlinecite{Kitaev1,Kwon, Fu1, Lutchyn, Law1}], resonant Andreev reflections [\onlinecite{Law2,Wimmer}], electron teleportations [\onlinecite{Bolech, Fu2}], as well as enhanced [\onlinecite{Nilsson}] and resonant [\onlinecite{James}] crossed Andreev reflections. Moreover, MFs in condensed matter systems obey non-Abelian statistics [\onlinecite{RG, Ivanov, Fujimoto, STF, Alicea2}] and have potential applications in fault-tolerant quantum computations [\onlinecite{Kitaev2, Nayak}]. 

In this work, we point out another intriguing phenomenon due to the self-Hermitian property of MFs, namely, MF-induced \emph{selective} equal spin Andreev reflections (SESARs). As depicted in Fig.1, when a spinful paramagnetic normal lead is coupled to a topological superconductor through its MF end state, electrons with spin pointing to a certain direction ${\bf \vec{n}}$ are reflected as holes with the same spin (Fig.1a), where ${\bf \vec{n}}$ is determined by the properties of the topological superconductor.    The reflected holes are created due to missing electrons with spin polarization $ \vec{\bf n}$ below the Fermi energy. Therefore, these processes are called equal spin Andreev reflections. This is in sharp contrast to ordinary Andreev reflection processes [\onlinecite{Blonder}], in which the reflected holes are created due to missing electrons below the Fermi energy which have opposite spins to the incoming electrons.

Even more interestingly, at the normal lead/topological superconductor (N/TS) junction,  electrons with opposite spin polarization $-\vec{\bf n}$ are completely \emph{decoupled} from the MF and they cannot participate in Andreev reflections (Fig.1b). In other words, the MF selects electrons with certain spin polarization $\vec{ \bf n}$ to undergo equal spin Andreev reflections. Therefore, we refer to this new phenomenon as MF-induced SESARs. 

\begin{figure}
\begin{center}
\includegraphics[width=3.2in]{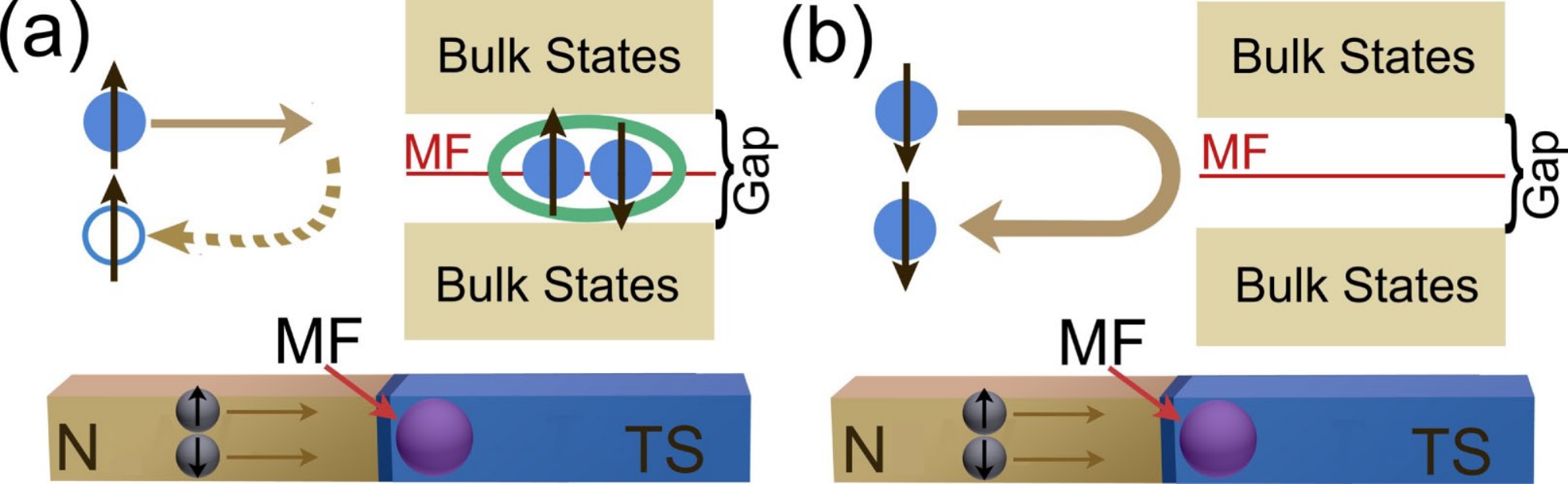}
\caption{ A paramagnetic normal lead (N) is coupled to a topological superconductor (TS) with MF end states. The zero energy MF mode is denoted by the horizontal line inside the bulk gap of the TS. (a) Electrons with a specific spin polarization can undergo equal spin Andreev reflections in which an electron is reflected as a hole with the same spin. (b) Electrons with opposite spin are totally reflected as electrons with unchanged spin. (c) Realizing a topological superconductor using a Rashba semi-conducting wire in proximity to an s-wave superconductor and in a magnetic field. The Rashba direction is denoted as $\vec{n}_{R}$. }
\label{Fig1}
\end{center}
\end{figure}

Pure equal spin Andreev reflections can take place at a half-metal/superconductor interface [\onlinecite{Eschrig1, Xiao, Visani, Eschrig2, Linder, Brouwer, Tanaka1, Niu}] if spin is not conserved at the interface. This is because conducting electrons in a half-metal are fully spin-polarized and usual Andreev reflection processes cannot occur. Nevertheless, as we show below, inducing SESARs in paramagnetic leads is a special property of MFs. Importantly, as in the half-metal case and depicted in Fig.1a, the charge current in the normal lead is fully spin-polarized since the current is carried by right-moving electrons and counter-propagating holes with the same spin. Therefore, a topological superconductor which supports MFs can be used as a novel device for inducing fully spin-polarized currents in paramagnetic leads. 

In the following sections, we first show, using an effective Hamiltonian approach, that SESARs are due to the self-Hermitian property of MFs. Second, we calculate the spin polarization direction ${\bf \vec{n}}$ of a N/TS junction. The topological superconductor is engineered by applying an external magnetic field to a semi-conducting wire in proximity to an s-wave superconductor [\onlinecite{Tewari, Alicea, ORV, PL11}] as depicted in Fig.1c. Third, we demonstrate how SESARs can be used to detect MFs in topological superconductors using a spin-polarized lead.

\emph{\bf Majorana-induced SESARs}--- At in-gap energy, the density of states at the ends of a topological superconducting wire is mainly due to zero energy MF end states. Therefore, we expect the transport properties of a N/TS junction at in-gap energy can be well described by an effective Hamiltonian which includes the lead and the coupling between the lead and the MF [\onlinecite{Bolech, Nilsson, Law1}]. The effective Hamiltonian $H_{T}$ can be written as:
\begin{equation}
\begin{array}{l}
H_{T} = H_{L} + H_{c}, \\
H_{L}  = -iv_{F} {\sum\limits_{\alpha \in {\uparrow / \downarrow} }} \int_{-\infty}^{+\infty}{\psi_{\alpha}^{\dag}(x)\partial_x \psi_{\alpha}(x)} {dx}, \\
H_{c}  = \tilde{t} \gamma [a \psi_{\uparrow}(0) + b \psi_{\downarrow}(0) -   a^{*} \psi_{\uparrow}^{\dagger}(0) - b^{*} \psi_{\downarrow}^{\dagger}(0)]. 
\end{array}
\end{equation}
Here, $H_{L}$ describes the normal lead with spin up and spin down electrons $\psi_{\uparrow/\downarrow}(x)$ and Fermi velocity $v_{F}$. The most general form of coupling between the MF end state $\gamma$ and the lead is described by $H_{c}$, where $\tilde{t}$ is a real number and $a$ and $b$ are complex numbers. The form of $H_{c}$ is determined by the self-Hermitian property of the MF $\gamma = \gamma^{\dagger}$ and the fact that $H_{c}$ is Hermitian. Without loss of generality, one can set $ |a|^2 + |b|^2 =1$. It is important to note that using a unitary transformation $ \Psi_{1}= a \psi_{\uparrow} + b \psi_{\downarrow}$ and $ \Psi_{2}= -b^{*} \psi_{\uparrow} + a^{*}\psi_{\downarrow}$, the Hamiltonian becomes 
\begin{equation}
\begin{array}{l}
H_{L}  = -iv_{F} {\sum\limits_{\alpha \in {1 / 2} }} \int_{-\infty}^{+\infty}{\Psi_{\alpha}^{\dag}(x)\partial_x \Psi_{\alpha}(x)} {dx}, \\
H_{c}  = \tilde{t} \gamma [\Psi_{1}(0) -  \Psi_{1}^{\dagger}(0)].
\end{array}
\end{equation}
Evidently, the MF only couples to the $\Psi_{1}$ electrons with spinor $\vec{s}_1 =|a| (1, b/a)^{T}= (\cos{\frac{\theta}{2}}, e^{i \phi} \sin{\frac{\theta}{2}})^{T}$, while the $\Psi_2$ electrons with spinor $\vec{s}_2 = (-\sin{\frac{\theta}{2}}, e^{i \phi} \cos{\frac{\theta}{2}})^{T}$ are totally decoupled from the MF. This Ising spin property of MFs [\onlinecite{Chung, Shindou, Simon, Flensberg, Choy}], which allows MFs to couple to electrons with certain spin polarization only, has significant effects on the transport properties of topological superconductors as we show below.

Since the $\Psi_2$ electrons are decoupled from the MF, we consider the $\Psi_{1}$ electrons and holes in the Hamiltonian in Eq.2 only. Denoting the incoming and outgoing electrons (holes) with energy $E$ relative to the Fermi energy as $\Psi_{1E}(-)$ ($\Psi_{1E}^{\dagger}(-)$ ) and $\Psi_{1E}(+)$ ($\Psi_{1E}^{\dagger}(+)$) respectively, the scattering matrix of the N/TS junction is:
\begin{equation}
\left( 
\begin{array}{cc}
\Psi_{1E} (+) \\
\Psi_{1E}^{\dagger}(+)
\end{array} \right) = 
\frac{1}{\Gamma+iE}\left( 
\begin{array}{cc}
{iE}  &   {\Gamma } \\
{\Gamma} & {iE}  \\
\end{array} \right)
\left( 
\begin{array}{cc}
\Psi_{1E} (-) \\
\Psi_{1E}^{\dagger}(-)
\end{array} \right), 
\end{equation}
where $\Gamma=2 \tilde{t}^2/v_F$. From the scattering matrix, we note that the $\Psi_1$ electrons are reflected as $\Psi_1$ holes with the same spin with Andreev reflection amplitude $\Gamma/(\Gamma+iE)$. From the spinors $\vec{s}_1$ and $\vec{s}_2$, we note that  $\Psi_1$ electrons have spins parallel to the direction ${\bf \vec{n}}=  \langle \vec{s}_1 | \vec{ \sigma} | \vec{s}_1 \rangle = (\sin \theta \cos \phi, \sin \theta \sin \phi, \cos \theta)$,  and $\Psi_2$ electrons have opposite spins, where $\vec{\bf \sigma}$ is the Pauli vector. Therefore, electrons with spin parallel to the ${\bf \vec{n}}$ directions can couple to the MF and undergo equal spin Andreev reflections, whereas electrons with opposite spin are totally reflected as electrons. We call this phenomenon MF-induced SESARs. 


\emph{\bf SESARs of spin-orbit coupled superconducting wires}--- The MF induced SESARs is a general phenomenon due to the self-Hermitian property of MFs as shown above. Moreover, $ \vec{\bf n}$ cannot be determined using the effective Hamiltonian. Therefore, to be specific, we study a N/TS junction where the topological superconductor can be realized experimentally [\onlinecite{kou, deng, das}] by applying a magnetic field to a spin-orbit coupled semi-conducting wire which is in proximity to an $s$-wave superconductor as depicted in Fig.1c.

In the Nambu basis $(\psi_{k \uparrow}, \psi_{k \downarrow}, \psi_{-k \uparrow}^{\dagger}, \psi_{-k \downarrow}^{\dagger})$, the Hamiltonian of the topological superconductor can be written as [\onlinecite{Tewari, Alicea, ORV, PL11}]:
\begin{equation}
H_{1D}(k)=[(\frac{k^2}{ 2m} -\mu) \sigma_0  + \vec{V} \cdot \vec{\sigma} + \alpha_{R} k \sigma_y) ]\tau_z - \Delta \sigma_y \tau_y.
\end{equation}
Here, $\psi_{k \uparrow}$ ($\psi_{k \downarrow}$) denotes a spin up (down) electron with momentum $k$, the effective mass and the chemical potential are denoted by $m$ and $\mu$ respectively. The Zeeman field is denoted by $\vec{V}$ and $\alpha_{R}$ is the Rashba spin-orbit coupling strength. The Pauli matrices $\sigma_{i}$ and $\tau_{i}$ act on the spin and particle-hole space respectively.

Suppose the one-dimensional superconducting wire occupies the semi-infinite space with $x \geq 0$ and a magnetic field with magnitude $V_z$ is applied along the $z$-direction, there exists a MF end state localized near $x=0$ in the topological regime when ${V_z}^2 > \mu^2 + \Delta^2$. The MF end state $\gamma$ satisfies the condition $H_{1D}(k \to -i\partial_{x}) \gamma = 0 $ with $\gamma^{\dagger} = \gamma$. In general, the Majorana mode can be written as:
\begin{equation} \label{MF}
\gamma (x) = \sum_{i=1}^{3} \beta_{i} \left( 
\begin{array}{c}
\vec{\phi}_{i}\\
\vec{\phi}_{i}
\end{array} \right) e^{- \lambda_{i}x } + \beta_{4} \left( 
\begin{array}{c}
\vec{\phi}_{4}\\
-\vec{\phi}_{4}
\end{array} \right) e^{- \lambda_{4}x },
\end{equation}
where $\lambda_i$ are the four solutions of the following two quartic equations with positive real parts
\begin{equation}
\left( \frac{\lambda^2}{2m} + \mu \right) ^2 + (\alpha_R \lambda \pm \Delta)^2 - V_{z}^2 =0.
\end{equation}

For realistic semi-conducting wires with $\frac{2m \alpha_R^2 } {\sqrt{V_z^2-\Delta^2}} \ll1 $ and at chemical potential $\mu  \approx 0$, we have $\lambda_1 = \lambda^{*}_{2}= i\lambda_0 + \delta $ and $\lambda_{3/4} = \lambda_{0} \mp \delta$, where $\lambda_0 = \sqrt{2m}\left(V_{z}^2-\Delta^2\right)^{1/4} $ and $\delta = 2 m ^2 \alpha_{R} \Delta/ \lambda_0^2$. Here, $\vec{\phi}_{i} = [ \lambda_{i}^2/(2m) + V_{z}, -\Delta - \alpha_{R} \lambda_{i}]^{T}$ for $\vec{\phi}_{1}$, $\vec{\phi}_{2}$ and $\vec{\phi}_{3}$, and $\vec{\phi}_{4} =  [\lambda_{4}^2/(2m) + V_{z}, \Delta - \alpha_{R} \lambda_{4}]^{T}$. 

Assuming that the lead can be described by the Hamiltonian $H_{L} = (k^2/2 m_{L} - \mu) \sigma_{0} \tau_{z} $, the wavefunction in the lead at the Fermi energy can be written as $\Psi_{L}(x)= \vec{e}_1 e^{ik_{F} x}+ d_{e \uparrow} \vec{e}_{1} e^{-i k_{F} x}+ d_{e \downarrow} \vec{e}_{2} e^{-i k_{F} x}+d_{h \uparrow} \vec{e}_{3}e^{i k_{F} x} + d_{h \downarrow} \vec{e}_{4}e^{i k_{F} x} $, where $k_{F}$ is the Fermi momentum and $\vec{e}_{1} = [1,0,0,0]^{T}$, $\vec{e}_{2} = [0,1,0,0]^{T}$, $\vec{e}_{3} = [0,0,1,0]^{T}$ and $\vec{e}_{4} = [0,0,0,1]^{T}$. Here, $d_{\alpha, \sigma}$ denotes the amplitude for an incoming spin up electron to be reflected as an electron ($e$) or hole ($h$) with spin $\sigma$. On the other hand, the wavefunction at the Fermi energy on the superconductor side $\Psi_{S}(x)$ can be written as the linear combination of the four-component vectors associated with $\vec{\phi_{i}}$ in Eq.{\ref{MF}}. We note that the wavefunction has to satisfy the continuity condition $\Psi_{L}(x)|_{x=0}=\Psi_{S}(x)|_{x=0}$ and current conservation condition $ J_{x} \Psi_{L}(x)|_{x=0} = J_{x} \Psi_{S}(x) |_{x=0}$, where the current operator is
\begin{equation}
J_{x}= \frac{\partial H_{1D}(k)}{\partial k} \vert_{k \to -i\partial_{x}} = 
\left(\begin{array}{cc}
-i \partial_{x}/m & -i \alpha_{R} \\
i \alpha_{R} & -i \partial_{x}/m
\end{array} \right)\tau_z.
\end{equation}
By solving the above boundary conditions, for both spin up and spin down incoming electrons, the scattering matrix of the N/TS junction at the Fermi energy can be found. At zeroth order in $\alpha_R$ with $\alpha_R \to 0$, the Andreev reflection matrix $r_{he}$, which relates the incoming electrons $(\psi_{k \uparrow}, \psi_{k \downarrow})^{T}$  with the outgoing holes $(\psi_{-k \uparrow}^{\dagger}, \psi_{-k \downarrow}^{\dagger})^{T}$, is:
\begin{equation} 
r_{he}(V_z) = \left(
\begin{array}{cc}
\frac{V_z - \sqrt{V_z^2-\Delta^2} }{2 V_z} & -\frac{\Delta}{2 V_z}\\
-\frac{\Delta}{2 V_z} & \frac{V_z + \sqrt{V_z^2-\Delta^2} }{2 V_z}\\
\end{array}
\right). \label{rhe}
\end{equation}

On the other hand, the normal reflection matrix which relates the incoming electrons with outgoing electrons is $r_{ee}(V_{z}) = r_{he}(-V_z) e^{i \chi(k)}$, where $e^{i\chi(k)} =\frac{ k/m_L-i \lambda_0/m}{ k/m_L+i  \lambda_0/m}$ is the phase acquired by the reflected electrons at the interface. Denoting $ \vec{s}_{0}= (\cos \frac{\theta_{0}}{2}, e^{i\phi_{0}}\sin \frac{\theta_{0}}{2})^{T} = \frac{1}{\mathcal{N}} (-\Delta, V_{z} + \sqrt{V_{z}^{2}-\Delta^2})^{T}$ with $\mathcal{N}$ the normalization factor, we have $r_{he}\vec{s}_0={\vec{s}_0}^{*}$ and $r_{ee} \vec{s}_0=0$. Therefore, to the zeroth order in $\alpha_R$, electrons in the conducting channel with spin parallel to ${\bf \vec{n}}_0=  \langle \vec{s}_0 |  \vec{ \sigma} | \vec{s}_0 \rangle = (\sin \theta_0 \cos \phi_0, \sin \theta_0 \sin \phi_0, \cos \theta_0)$ will be resonantly reflected as holes with the same spin. On the contrary, electrons with spinor $ \vec{u}_{0}= (-\sin \frac{\theta_{0}}{2}, e^{i\phi_{0}}\cos \frac{\theta_{0}}{2})$ and spin anti-parallel to ${\bf \vec{n}}_0$ are totally reflected as electrons with unchanged spin since $r_{ee}\vec{u}_0=e^{i \chi}\vec{u}_0$. 

It is important to note that the form of $r_{he}$ strongly depends on the existence of the MF solution in Eq.5. In the trivial regime, $r_{he}$ will be dominated by off-diagonal elements and usual opposite spin Andreev reflection processes will dominate. It is also interesting to note that, in the weak coupling regime and weak Rashba, the electronic part of the MF wavefunction in Eq.5 is approximately proportional to $\vec{s}_{0}$. Therefore, the measurement of $\vec{\bf n}$ can reveal the form of the MF wavefunction.

\begin{figure}
\begin{center}
\includegraphics[width=3.2in]{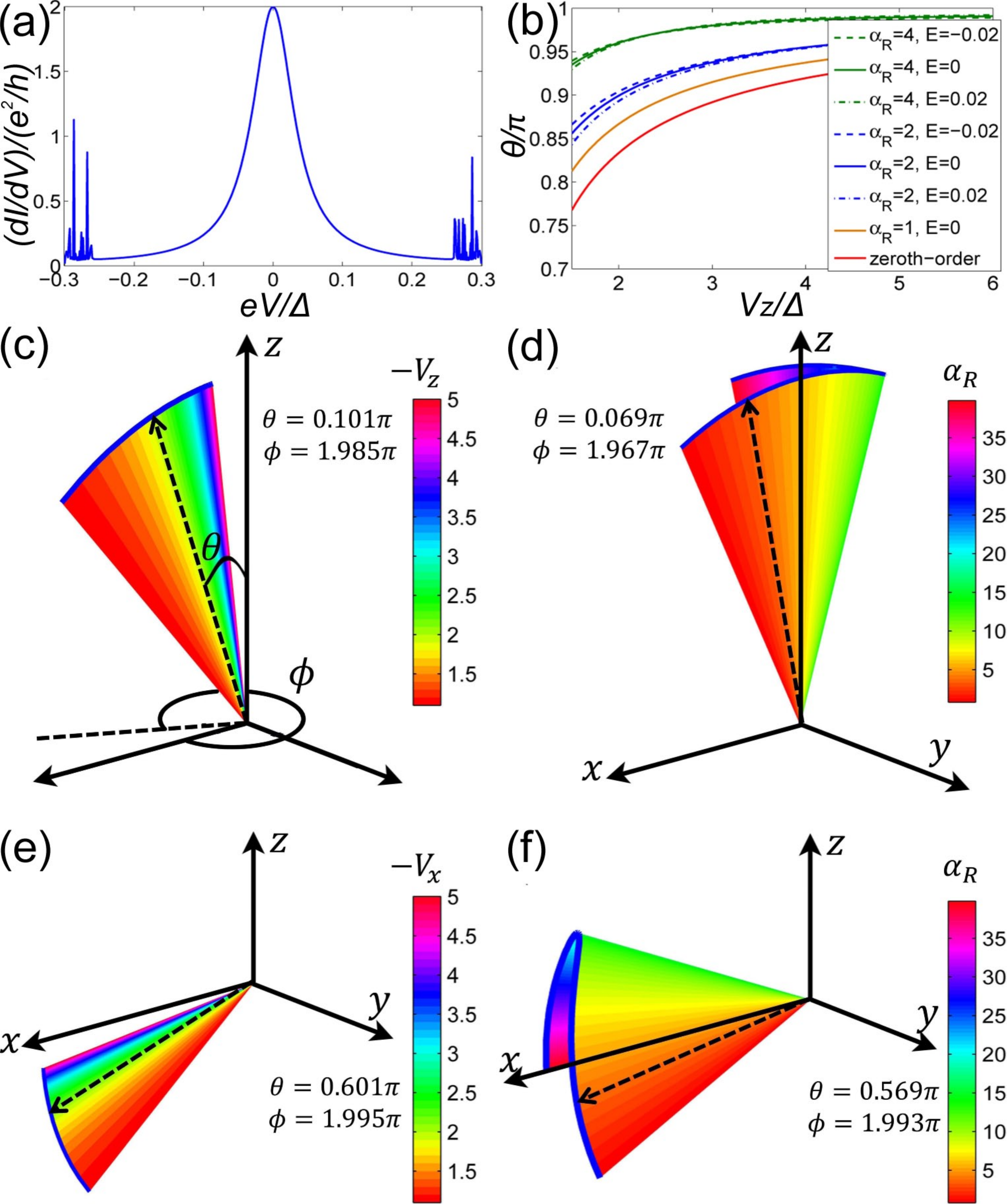}
\caption{ $\Delta=1$, $t=25$, $t'=30$, $t_c=15$ for all the figures. (a) The differential conductance $dI/dV$ of the N/TS junction as a function of voltage bias $V$. The parameters are chosen as: $\alpha_R=2$, $V_z=2$. (b) The angle $\theta$ of the polarization vector ${ \bf \vec{n}} $  as a function of $V_z$, for different $\alpha_R$ and voltage bias. The zeroth order result from ${\bf \vec{n}}_0$, which is a good approximation for the numerical results for small $\alpha_R$, is also presented.  (c)-(f) The polarization vector ${\bf \vec{n}}$ for different parameters at zero voltage bias. The coordinates $\theta$ and $\phi$ denote the coordinates of the dashed vector. (c) ${\bf \vec{n}}$ with $\alpha_R=2$ and different $V_z$. $V_z=-2$ for the dashed vector. (d) ${\bf \vec{n}}$ with $V_z=-2$ at different $\alpha_R$. $\alpha_R=3$ for the dashed vector. (e) ${\bf \vec{n}}$ with $\alpha_R=2$ and different $V_x$. $V_x=-2$ for the dashed vector (f) ${\bf \vec{n}}$ with $V_x=-2$ and different $\alpha_R$. $\alpha_R=3$ for the dashed vector. }
\label{Fig1}
\end{center}
\end{figure}

To further verify the analytic results and generalize the results to arbitrary Rashba strength and voltage bias, we calculate the scattering matrix of the N/TS junction using a tight-binding model used in Refs.[\onlinecite{Jie2, Jie}].

The scattering matrix of the N/TS junction can be calculated using the recursive Green's function method [\onlinecite{Lee,Jie}]. For example, the reflection matrix elements for an incoming electron are:
\begin{equation}
\tilde{r}^{\sigma'\sigma}_{\alpha e}= -\delta_{\sigma \sigma'} \delta_{\alpha e} + i[ \Gamma^{1/2}]_{\sigma'}^{\alpha} *[G^r]_{\alpha e}^{\sigma' \sigma}*[ \Gamma^{1/2}]_{\sigma}^{e}.
\end{equation}
Here, $\tilde{r}_{\alpha e}^{\sigma' \sigma}$ is the reflection amplitude of an incoming electron with spin $\sigma$ to be reflected as an $\alpha$ particle with spin $\sigma'$ where $\alpha$ denotes electron ($e$) or hole ($h$). $[G^r]_{\alpha e}^{\sigma' \sigma}$ is a matrix element of the retarded Green's function $G^r$ of the superconductor. The broadening function is denoted by $\Gamma_{\sigma}^{\alpha}= i[(\Sigma_{\sigma}^{\alpha})^r- (\Sigma_{\sigma}^{\alpha})^a]$, where $(\Sigma_{\sigma}^{\alpha})^{r(a)}$ is the retarded (advanced) self-energy of the $\alpha$ particle lead with spin $\sigma$.

Numerically we find that, in the topological regime, there are two eigenvectors $\vec{s}_{n}$ and $\vec{u}_{n}$ for the normal reflection matrix $\tilde{r}_{ee}$ with $\tilde{r}_{ee} \vec{s}_{n} = m_{1} \vec{s}_{n}$ and $\tilde{r}_{ee} \vec{u}_{n} = m_{2} \vec{u}_{n}$ respectively. For the Andreev reflection matrix, we have $\tilde{r}_{he}\vec{s}_{n} = m'_{1} \vec{s}_{n}^{*}$ and $\tilde{r}_{he}\vec{u}_{n} = 0$.  The eigenvalues are in general complex and have the properties $|m_{1}|<1$, $|m_2| =1$ and $|m'_{1}| \leq1$.  This shows that electrons with spinor $\vec{u}_{n}$ are reflected as electrons with the same spin with probability of unity. On the other hand, electrons with spinor $\vec{s}_{n}$ can be reflected as holes with the same spin with Andreev reflection amplitude $m'_{1}$.  In other words, electrons with spin polarization ${\bf \vec{n}}= \langle \vec{s}_n|  \vec{ \sigma} | \vec{s}_n \rangle$ can undergo equal spin Andreev reflections. Electrons with opposite spin polarization  $-{\bf \vec{n}}= \langle \vec{u}_n|  \vec{ \sigma} | \vec{u}_n \rangle$ are totally reflected. This is consistent with the effective Hamiltonian and the analytic results.

The differential conductance $dI/dV$, as a function of voltage bias $V$ between the lead and the superconductor, is shown in Fig.2a. As expected, the zero bias conductance is quantized to $2e^2/h$ as the MF couples to only a single conducting channel of the lead.

To study the spin polarization vector ${\bf \vec{n}}= \langle \vec{s}_n|  \vec{ \sigma} | \vec{s}_n \rangle$, we plot the angle $\theta$ calculated from the tight-binding model [\onlinecite{Jie, Jie2}] as a function of $V_{z}$ for different incoming electron energy $eV$ and different $\alpha_R$. The results are shown in Fig.2b. The zeroth order analytic result at zero bias, which is a good approximation for the numerical results for small $\alpha_R$, is also shown in Fig.2b. The finite voltage bias results are denoted by dashed lines. It is important to note that $\theta$ is not sensitive to the energy of the incoming electrons so that the current at finite bias is also spin-polarized.

In Fig.2c, ${\bf \vec{n}}$ as a function of $V_z$ is depicted. As expected from the analytic results for $\alpha_R \to 0$ that $\vec{\bf n}_{0} = \langle \vec{s}_0|  \vec{ \sigma} | \vec{s}_0 \rangle $ with $ \vec{s}_{0} \propto (-V_{z} + \sqrt{V_{z}^{2}-\Delta^2}, \Delta)^{T} $, the projection of ${\bf \vec{n}}$ on the $z$-axis increases as $|V_z|$ increases. On the other hand, $\phi=0$ when $\alpha_R \to 0$ as the Andreev reflection matrix in Eq.8 is real.  For small Rashba strength, $\phi$ is only weakly dependent on $V_z$ and it deviates only slightly from $2 \pi$. The ${\bf \vec{n}}$ dependent on $\alpha_R$ for fixed $V_z$ is shown in Fig.2d. Experimentally, it is also convenient to apply the magnetic field along the wire so that $V_x$ is finite. For $\alpha_R \to 0$, the polarization vector is $\vec{\bf n}_{0} = \langle \vec{s}_0|  \vec{ \sigma} | \vec{s}_0 \rangle $ with $\vec{s}_{0} \propto (\sqrt{V_x^2-\Delta^2},\Delta-V_x)^T $. The numerical results for the $V_x$ and $\alpha_R$ dependence of ${\bf \vec{n}}$ are shown in Figs.2e and 2f respectively. 

\emph{\bf Coupling between MFs and spin-polarized leads}--- It is shown above that incoming electrons with different spin polarizations interact with the topological superconductor differently. Electrons with spin parallel to ${\bf \vec{n}}$ can undergo equal spin Andreev reflections, whereas electrons with opposite spin polarization are totally reflected as electrons. Therefore, if the normal lead is spin-polarized, we expect that the conductance of a N/TS junction will strongly depend on the spin polarization of the lead. 

\begin{figure}
\begin{center}
\includegraphics[width=3.2in]{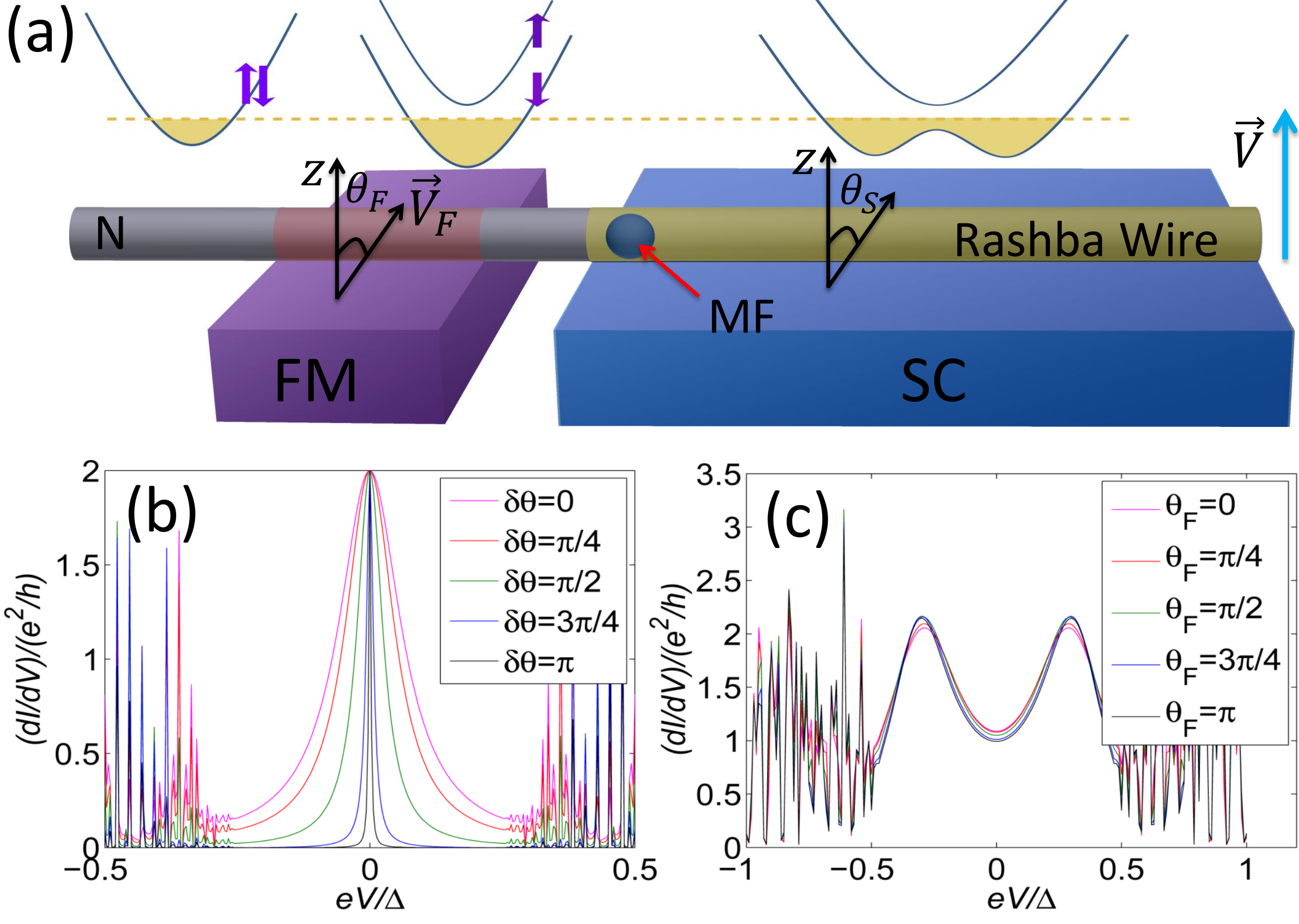}
\caption{ (a) A normal lead (N) is coupled to a semi-conducting wire with Rashba spin-orbit coupling and in proximity to a superconductor (SC). The wire can support MF end states. A ferromagnetic (FM) section is added to the normal lead to polarize the electrons of the wire. The schematic band structure of different sections of the wire are shown. The Fermi energy is denoted by the yellow dashed line. The spin degeneracy of the spin up and spin down bands in the ferromagnetic section of the normal lead is lifted. (b) The differential conductance as a function of $\delta \theta$ in the topological regime with MFs. The tight-binding parameters are the same as in Fig.2a except that a Zeeman field $\vec{V}_F$ with $|\vec{V}_{F}|=10\Delta$ is applied to a section of 20 sites of the normal lead, which is 10 sites away from the N/TS interface. (c) The differential conductance as a function of $\theta_F$ in the topologically trivial regime.}
\label{Fig1}
\end{center}
\end{figure}

The experimental setup is depicted in Fig.3a in which a normal lead is coupled to one end of a topological superconductor. A ferromagnet is coupled to a section of the normal lead so that electrons passing through the magnetic section is strongly polarized by the ferromagnet. The schematic band structure of different sections of the system is shown in Fig.3a. By controlling the magnetization direction of the ferromagnet, one can control the spin polarization direction of the incoming electrons at the N/TS junction.

We denote the polarization angle of the ferromagnet and the topological superconductor with respect to the $z$-axis as $\theta_F$ and $\theta_S$ respectively. The conductance of the N/TS junction for different angles $\delta \theta =\theta_F-\theta_S$ is shown in Fig.3b. When $\delta \theta \approx 0$, most of the incoming electrons can undergo equal spin Andreev reflections. As a result, the width of the conductance peak, which measures the coupling strength between the lead and the topological superconductor, is wide. As $\delta \theta$ deviates from zero, the incoming electrons can be decomposed into the Andreev reflected channel and the totally reflected channel. As $\delta \theta$ increases, the weight of the totally reflected channel becomes more important and the width of the conductance peak becomes narrower. Nevertheless, the height of the zero bias conductance peak at zero temperature is not changed due to resonant Andreev reflections. On the contrary, in the topologically trivial regime, in which two transverse subbands of the semiconductor wire are occupied, Andreev reflections are mainly induced by ordinary fermionic end states and ordinary Andreev reflection processes will dominate. As a result, the conductance is only weakly dependent on $\theta_F$. Therefore, the MF-induced SESARs can be used to distinguish the topological regime from the trivial regime of the superconductor.

\emph{\bf Conclusion}--- In short, we show in this work that MFs induce SESARs. As a result, topological superconductors can be used as novel devices to generate spin-polarized currents in paramagnetic leads. The SESARs can also be used to detect MFs if spin-polarized leads are used.

\emph{Acknowledgments}--- We thank Chris LM Wong for discussions. JJH and KTL thank the support from HKRGC through Grants 605512 and 602813. PAL acknowledges the support by DOE Grant DE-FG-02-03-ER46076.


\begin{thebibliography}{99}
\bibitem{Wilczek} F. Wilczek, Nat. Phys. {\bf 5}, 614 (2009).
\bibitem{Kitaev1} A. Y. Kitaev, Physics-Uspekhi {\bf44}, 131 (2001).
\bibitem{Kwon} H.J. Kwon, K. Sengupta and V.M. Yakovenko, Eur. Phys. J. B {\bf37}, 349 (2004).
\bibitem{Fu1} L. Fu and C. Kane, Phys. Rev. B {\bf 79}, 161408(R) (2009).
\bibitem{Lutchyn} R. M. Lutchyn, J. D. Sau, and S. Das Sarma, Phys. Rev. Lett.{\bf105}, 077001 (2010).
\bibitem{Law1}K. T. Law and P. A. Lee, Phys. Rev. B {\bf84}, 081304 (R) (2011).
\bibitem{Law2} K. T. Law, P. A. Lee, and T. K. Ng, Phys. Rev. Lett. {\bf 103}, 237001 (2009).
\bibitem{Wimmer} M. Wimmer, A.R. Akhmerov, J.P. Dahlhaus, C.W.J. Beenakker, New J. Phys. {\bf13}, 053016 (2011).
\bibitem{Bolech} C. J. Bolech and E. Demler, Phys. Rev. Lett. {\bf98}, 237002 (2007).
\bibitem{Fu2} L. Fu, Phys. Rev. Lett. {\bf 104}, 056402 (2010).
\bibitem{Nilsson} J. Nilsson, A. R. Akhmerov, C. W. J. Beenakker, Phys. Rev. Lett. {\bf 101},120403 (2008).
\bibitem{James} J. J. He, J. Wu, T-P Choy, X-J Liu, Y. Tanaka, K. T. Law, arXiv:1307.2764 (To appear in Nat. Comm.).
\bibitem{RG} N. Read and D. Green, Phys. Rev. B {\bf61}, 10267 (2000).
\bibitem{Ivanov} D. A. Ivanov, Phys. Rev. Lett. {\bf86}, 268 (2001).
\bibitem{Fujimoto} S. Fujimoto, Phys. Rev. B {\bf77}, 220501 (2008).
\bibitem{STF} M.Sato, Y. Takahashi, S. Fujimoto, Phys. Rev. Lett. {\bf103} 020401 (2009).
\bibitem{Alicea2} J. Alicea, Y. Oreg, G. Refael, F. von Oppen, M. P. A. Fisher Nat. Phys. {\bf7},1915 (2011).
\bibitem{Kitaev2} A. Kitaev, Ann. Phys. {\bf 303}, 2-30 (2003).
\bibitem{Nayak} C. Nayak, S. H. Simon, A. Stern, M. S. Freedman and S. Das Sarma,  Rev. Mod. Phys. {\bf 80}, 1083-1159 (2008).
\bibitem{Blonder} G. E. Blonder, M. Tinkham and T. M. Klapwijk, Phys. Rev. B, {\bf25} 4515 (1982).
\bibitem{Eschrig1} M. Eschrig, J. Kopu, J. C. Cuevas, and G. Schon, Phys. Rev. Lett. {\bf90}, 137003 (2003).
\bibitem{Xiao} R. S. Keizer, S. T. B. Goennenwein, T. M. Klapwijk, G. Miao, G. Xiao and A. Gupta, Nature 439, 825�827 (2006).
\bibitem{Visani}  C. Visani, Z. Sefrioui, J. Tornos, C. Leon, J. Briatico, M. Bibes, A. Barthelemy, J. Santamaria and J. E. Villegas, Nat. Phys. {\bf8}, 539�543 (2012).
\bibitem{Eschrig2} M. Eschrig and T. Lofwander, Nat. Phys. {\bf4}, 138 (2008).
\bibitem{Linder} J. Linder, M. Cuoco and A. Sudbo, Phys. Rev. B {\bf 81}, 174526 (2010).
\bibitem{Brouwer} M. Duckheim and P. W. Brouwer, Phys. Rev. B {\bf 83}, 054513 (2011).
\bibitem{Tanaka1} Y. Tanaka, M. Sato, N. Nagaosa, J. Phys. Soc. Jpn. {\bf 81}, 011013 (2012).
\bibitem{Niu} Z. P. Niu,  Euro. Phys. Lett. {\bf100} 17012 (2012).
\bibitem{Tewari} J. Sau, R. M. Lutchyn, S. Tewari and S. Das Sarma, Phys. Rev. Lett. {104}, 040502 (2010).
\bibitem{Alicea} J. Alicea, Phys. Rev. B {\bf 81}, 125318 (2010).
\bibitem{ORV} Y. Oreg, G. Refael, and F. von Oppen, Phys. Rev. Lett. {\bf105}, 177002 (2010).
\bibitem{PL11} A.C. Potter, P.A. Lee, Phys. Rev. B {\bf 83}, 094525 (2011).
\bibitem{Chung} S. B. Chung and S. C. Zhang, Phys. Rev. Lett. {\bf 103}, 235301 (2009).
\bibitem{Shindou} R. Shindou, A. Furusaki, N. Nagaosa, Phys. Rev. B {\bf 82}180505, (2010).
\bibitem{Simon} D. Sticlet, C. Bena and P. Simon, Phys. Rev. Lett. {\bf108}, 096802 (2012).
\bibitem{Flensberg} M. Leijnse and K. Flensberg, Phys. Rev. Lett.  {\bf107}, 210502 (2011).
\bibitem{Choy} T. P. Choy, K.T. Law and T. K. Ng, arXiv:1301.2068 (2013).
\bibitem{kou} V. Mourik, K. Zuo, S.M. Frolov, S.R. Plissard, E.P.A.M. Bakkers, L.P. Kouwenhoven, Science {\bf 336}, 1003 (2012).
\bibitem{deng} M. T. Deng, C.L. Yu, G.Y. Huang, M. Larsson, P. Caroff, H.Q. Xu, Nano Lett. {\bf12}, 6414-6419 (2012).
\bibitem{das} A. Das, Y. Ronen, Y. Most, Y. Oreg, M. Heiblum, H.Shtrikman, Nature Physics {\bf 8}, 887 (2012).
\bibitem{Jie2} J. Liu, A. C. Potter, K.T. Law and  P. A. Lee, Phys. Rev. Lett. 109, 267002 (2012).
\bibitem{Jie} J. Liu, F-C Zhang and  K. T. Law  Phys. Rev. B {\bf88} , 064509 (2013).
\bibitem{Lee} P. A. Lee and D. S. Fisher, Phys. Rev. Lett. {\bf 47}, 882 (1981).
\end{thebibliography}
\end{document}